\documentclass[aps,twocolumn,superscriptaddress]{revtex4}

\usepackage{graphicx}

\begin{document}

%\begin{center}

\title{Rabi oscillations in a qubit coupled to a quantum two-level system}

\author{S. Ashhab}
\affiliation{Frontier Research System, The Institute of Physical
and Chemical Research (RIKEN), Wako-shi, Saitama 351-0198, Japan}
\author{J. R. Johansson}
\affiliation{Frontier Research System, The Institute of Physical
and Chemical Research (RIKEN), Wako-shi, Saitama 351-0198, Japan}
\author{Franco Nori}
\affiliation{Frontier Research System, The Institute of Physical
and Chemical Research (RIKEN), Wako-shi, Saitama 351-0198, Japan}
\affiliation{Center for Theoretical Physics, CSCS, Department of
Physics, University of Michigan, Ann Arbor, Michigan 48109-1040,
USA}

\date{\today}

\begin{abstract}
We consider the problem of a qubit driven by a harmonically
oscillating external field while it is coupled to a quantum
two-level system. We perform a systematic numerical analysis of
the problem by varying the relevant parameters. The numerical
calculations agree with the predictions of a simple intuitive
picture, namely one that takes into consideration the four-level
energy spectrum, the simple principles of Rabi oscillations and
the basic effects of decoherence. Furthermore, they reveal a
number of other interesting phenomena. We provide explanations for
the various features that we observe in the numerical calculations
and discuss how they can be used in experiment. In particular, we
suggest an experimental procedure to characterize an environment
of two-level systems.
\end{abstract}

\maketitle

\section{Introduction}

There have been remarkable advances in the field of
superconductor-based quantum information processing in recent
years \cite{You1}. Coherent oscillations and basic gate operations
have been observed in systems of single qubits and two interacting
qubits
\cite{Nakamura1,Nakamura2,Vion,Yu,Martinis1,Chiorescu1,Pashkin,Yamamoto}.
One of the most important operations that are used in manipulating
qubits is the application of an oscillating external field on
resonance with the qubit to drive Rabi oscillations
\cite{Nakamura2,Vion,Yu,Martinis1,Simmonds}. A closely related
problem with great promise of possible applications is that of a
qubit coupled to a quantum harmonic-oscillator mode
\cite{You2,Chiorescu2,Wallraff,Johansson}.

Qubits are always coupled to uncontrollable degrees of freedom
that cause decoherence in its dynamics. One generally thinks of
the environment as slowly reducing the coherence of the qubit,
typically as a monotonically decreasing decay function. In some
recent experiments, however, oscillations in the qubit have been
observed that imply it is strongly coupled to quantum degrees of
freedom with long decoherence times \cite{Simmonds,Cooper}. The
effects of those degrees of freedom have been successfully
described by modelling them as quantum two-level systems (TLSs)
\cite{Ku,Shnirman,Faoro,Galperin,Ashhab}. Since, as mentioned
above, Rabi oscillations are a simple and powerful method to
manipulate the quantum state of a qubit, it is important to
understand the behaviour of a qubit that is driven on or close to
resonance in the presence of such a TLS. Furthermore, we shall
show below that driving the qubit close to resonance can be used
to extract more parameters about an environment of TLSs than has
been done in experiment so far. The results of this study are also
relevant to the problem of Rabi oscillations in a qubit that is
interacting with other surrounding qubits.

Some theoretical treatments and analysis of special cases of the
problem at hand were given in Refs. \cite{Cooper,Galperin}. In
this paper we perform a more systematic analysis in order to reach
a more complete understanding of this phenomenon. We shall present
a few simple physical principles that can be used to understand
several aspects of the behaviour of this system with different
possible choices of the relevant parameters. Those principles are
(1) the four-level energy spectrum of the qubit+TLS system, (2)
the basic properties of the Rabi-oscillation dynamics and (3) the
basic effects of decoherence. We shall then perform numerical
calculations that will agree with that intuitive picture and also
will reveal other results that are more difficult to definitively
predict otherwise. Finally, we suggest an experimental procedure
where the driven qubit dynamics can be used to characterize the
environment of TLSs.

The paper is organized as follows: in Sec. II we introduce the
model system and the Hamiltonian that describes it. In Sec. III we
present a few simple arguments that will be used as a foundation
for our numerical analysis of Sec. IV, which will confirm that
intuitive picture and reveal other less intuitively predictable
results (note that a reader who is sufficiently familiar with the
subject matter can skip Sec. III). In Sec. V we discuss how our
results can be used in experiment. We finally conclude our
discussion in Sec. VI.

\section{Model system}

The model system that we shall study in this paper is composed of
a harmonically driven qubit, a quantum TLS and their weakly
coupled environment \cite{Assumptions}. We assume that the qubit
and the TLS interact with their own (uncorrelated) environments
that would cause decoherence even in the absence of qubit-TLS
coupling. The Hamiltonian of the system is given by:

\begin{equation}
\hat{H}(t) = \hat{H}_{\rm q}(t) + \hat{H}_{\rm TLS} + \hat{H}_{\rm
I} + \hat{H}_{\rm Env},
\end{equation}

\noindent where $\hat{H}_{\rm q}$ and $\hat{H}_{\rm TLS}$ are the
qubit and TLS Hamiltonians, respectively; $\hat{H}_{\rm I}$
describes the coupling between the qubit and the TLS, and
$\hat{H}_{\rm Env}$ describes all the degrees of freedom in the
environment and their coupling to the qubit and TLS. The
(time-dependent) qubit Hamiltonian is given by:

\begin{eqnarray}
\hat{H}_{\rm q}(t) = &-& \frac{\Delta_{\rm q}}{2}
\hat{\sigma}_x^{\rm (q)} - \frac{\epsilon_{\rm q}}{2}
\hat{\sigma}_z^{\rm (q)}\\\nonumber
 &+& F \cos(\omega t) \left(
\sin \theta_{\rm f} \hat{\sigma}_x^{\rm (q)} + \cos \theta_{\rm f}
\hat{\sigma}_z^{\rm (q)} \right), \label{eq:QubitHamiltonian}
\end{eqnarray}

\noindent where $\Delta_{\rm q}$ and $\epsilon_{\rm q}$ are the
adjustable static control parameters of the qubit,
$\hat{\sigma}_{\alpha}^{\rm (q)}$ are the Pauli spin matrices of
the qubit, $F$ and $\omega$ are the amplitude (in energy units)
and frequency, respectively, of the driving field, and
$\theta_{\rm f}$ is an angle that describes the orientation of the
external field relative to the qubit $\hat\sigma_z$ axis. Although
we have used a rather general form to describe the coupling
between the qubit and the driving field, we shall see in Sec. III
that one only needs a single, easily extractable parameter to
characterize the amplitude of the driving field. We assume that
the TLS is not coupled to the external driving field, and its
Hamiltonian is given by:

\begin{equation}
\hat{H}_{\rm TLS} = - \frac{\Delta_{\rm TLS}}{2}
\,\hat{\sigma}_x^{\rm (TLS)} - \frac{\epsilon_{\rm TLS}}{2}
\,\hat{\sigma}_z^{\rm (TLS)},
\end{equation}

\noindent where the definition of the parameters and operators is
similar to those of the qubit, except that the TLS parameters are
uncontrollable. Note that our assumption that the TLS is not
coupled to the driving field can be valid even in cases where the
physical nature of the TLS and the driving field leads to such
coupling, since we generally consider a microscopic TLS, rendering
any coupling to the external field negligible.

The energy splitting between the two quantum states of each
subsystem, in the absence of coupling between them, is given by:

\begin{equation}
E_{\alpha} = \sqrt{{\Delta_{\alpha}}^2+{\epsilon_{\alpha}}^2},
\end{equation}

\noindent where the index $\alpha$ refers to either the qubit or
the TLS. The corresponding ground and excited states are,
respectively, given by:

\begin{eqnarray}
|g \rangle_{\alpha} = \cos \frac{\theta_{\alpha}}{2} | \uparrow
\rangle_{\alpha} + \sin \frac{\theta_{\alpha}}{2} | \downarrow
\rangle_{\alpha} \nonumber \\
|e \rangle_{\alpha} = \sin \frac{\theta_{\alpha}}{2} | \uparrow
\rangle_{\alpha} - \cos \frac{\theta_{\alpha}}{2} | \downarrow
\rangle_{\alpha},
\end{eqnarray}

\noindent where the angle $\theta_{\alpha}$ is given by the
criterion $\tan \theta_{\alpha} =
\Delta_{\alpha}/\epsilon_{\alpha}$. We take the interaction
Hamiltonian between the qubit and the TLS to be of the form:

\begin{equation}
\hat{H}_{\rm I} = - \frac{\lambda}{2} \,\hat{\sigma}_z^{\rm (q)}
\otimes \hat{\sigma}_z^{\rm (TLS)},
\end{equation}

\noindent where $\lambda$ is the (uncontrollable) coupling
strength between the qubit and the TLS. Note that any interaction
Hamiltonian that is a product of a qubit observable times a TLS
observable can be recast into the above form using a simple basis
transformation, keeping in mind that such a basis transformation
also changes the values of $\theta_{\rm q}$, $\theta_{\rm TLS}$
and $\theta_{\rm f}$.

We assume that all the coupling terms in $\hat{H}_{\rm Env}$ are
small enough that its effect on the dynamics of the qubit+TLS
system can be treated within the framework of the markovian
Bloch-Redfield master equation approach. We shall use a noise
power spectrum that can describe both dephasing and relaxation
with independently adjustable rates, and shall present our results
in term of those decoherence rates. For definiteness in the
numerical calculations, we take the coupling of the qubit and the
TLS to their respective environments to be described by the
operators $\hat{\sigma}_z^{(\alpha)}$, where $\alpha$ refers to
the qubit and the TLS. Note, however, that since we use the
relaxation and dephasing rates to quantify decoherence, our
results are independent of the choice of system-environment
coupling operators.

\section{Intuitive picture}

We start our analysis of the problem by presenting a few physical
principles that prove very helpful in intuitively predicting the
behaviour of the above-described system. Note that the arguments
given in this section are well known
\cite{Baym,Cohen_Tannoudji,Slichter}. For the sake of clarity,
however, we present them explicitly and discuss their roles in the
problem at hand.

\subsection{Energy levels and eigenstates}

The first element that one needs to consider is the energy levels
of the combined qubit+TLS system. In order for a given
experimental sample to function as a qubit, the qubit-TLS coupling
strength $\lambda$ must be much smaller than the energy splitting
of the qubit $E_{\rm q}$. We therefore take that limit, as well as
the limit $\lambda \ll E_{\rm TLS}$, and straightforwardly find
the energy levels to be given by:

%\begin{widetext}
\begin{eqnarray}
E_1 & = & - \frac{E_{\rm TLS}+E_{\rm q}}{2} - \frac{\lambda_{\rm
cc}}{2} \nonumber
\\
E_2 & = & - \frac{1}{2} \sqrt{ \left(E_{\rm TLS}-E_{\rm
q}\right)^2 + \lambda_{\rm ss}^2} + \frac{\lambda_{\rm cc}}{2}
\nonumber
\\
E_3 & = & + \frac{1}{2} \sqrt{ \left(E_{\rm TLS}-E_{\rm
q}\right)^2 + \lambda_{\rm ss}^2} + \frac{\lambda_{\rm cc}}{2}
\nonumber
\\
E_4 & = & + \frac{E_{\rm TLS}+E_{\rm q}}{2} - \frac{\lambda_{\rm
cc}}{2},
\end{eqnarray}
%\end{widetext}

\noindent where $\lambda_{\rm cc}= \lambda \cos\theta_{\rm
q}\cos\theta_{\rm TLS}$, $\lambda_{\rm ss}= \lambda
\sin\theta_{\rm q}\sin\theta_{\rm TLS}$. The corresponding
eigenstates are given by:

\begin{eqnarray}
| 1 \rangle & = & | gg \rangle \nonumber
\\
| 2 \rangle & = & \cos \frac{\varphi}{2} | eg \rangle + \sin
\frac{\varphi}{2} | ge \rangle \nonumber
\\
| 3 \rangle & = & \sin \frac{\varphi}{2} | eg \rangle - \cos
\frac{\varphi}{2} | ge \rangle \nonumber
\\
| 4 \rangle & = & | ee \rangle,
\end{eqnarray}

\noindent where the first symbol refers to the qubit state and the
second one refers to the TLS state in their respective uncoupled
bases, the angle $\varphi$ is given by the criterion $\tan \varphi
= \lambda_{\rm ss} / \left(E_{\rm TLS}-E_{\rm q}\right)$, and for
definiteness in the form of the states $| 2 \rangle$ and $| 3
\rangle$ we have assumed that $E_{\rm TLS} \geq E_{\rm q}$.

Note that the mean-field shift of the qubit resonance frequency,
$\lambda_{\rm cc}$, is present regardless of the values of the
qubit and TLS energy splittings. The avoided-crossing structure
involving states $|2\rangle$ and $|3\rangle$, however, is only
relevant when the qubit and TLS energies are almost equal. One can
therefore use spectroscopy of the four-level structure to
experimentally measure the TLS energy splitting $E_{\rm TLS}$ and
angle $\theta_{\rm TLS}$, as will be discussed in more detail in
Sec. V.

\subsection{Rabi oscillations}

If a two-level system (e.g. a qubit) with energy splitting
$\omega_0$, initially in its ground state, is driven by a
harmonically-oscillating weak field with a frequency $\omega$
close to its energy splitting (up to a factor of $\hbar$) as
described by Eq. (\ref{eq:QubitHamiltonian}), its probability to
be found in the excited state at a later time $t$ is given by:

\begin{equation}
P_e = \frac{\Omega_0^2}{\Omega_0^2+(\omega-\omega_0)^2} \frac{1 -
\cos (\Omega t)}{2}
\end{equation}

\noindent where $\Omega=\sqrt{\Omega_0^2+(\omega-\omega_0)^2}$,
and the on-resonance Rabi frequency $\Omega_0= F |\sin
(\theta_{\rm f}-\theta_{\rm q})|$ (we take $\hbar=1$)
\cite{Slichter}. We therefore see that maximum oscillations with
full $g \leftrightarrow e$ conversion probability are obtained
when the driving is resonant with the qubit energy splitting. We
also see that the width of the Rabi peak in the frequency domain
is given by $\Omega_0$. Simple Rabi oscillations can also be
observed in a multi-level system if the driving frequency is on
resonance with one of the relevant energy splittings but off
resonance with all others.

\subsection{The effect of decoherence}

In an undriven system, the effect of decoherence is to push the
density matrix describing the system towards its
thermal-equilibrium value with time scales given by the
characteristic dephasing and relaxation times. The effects of
decoherence, especially dephasing, can be thought of in terms of a
broadening of the energy levels. In particular, if the energy
separation between the states $| 2 \rangle$ and $| 3 \rangle$ of
Sec. III-A is smaller than the typical decoherence rates in the
problem, any effect related to that energy separation becomes
unobservable. Alternatively, one could say that only processes
that occur on a time scale faster than the decoherence times can
be observed.

It is worth taking a moment to look in some more detail at the
problem of a resonantly-driven qubit coupled to a dissipative
environment, which is usually studied under the name of
Bloch-equations \cite{Slichter,Smirnov}. If the Rabi frequency is
much smaller than the decoherence rates, the qubit will remain in
its thermal equilibrium state, since any deviations from that
state caused by the driving field will be dissipated immediately.
If, on the other hand, the Rabi frequency is much larger than the
decoherence rates, the system will perform damped Rabi
oscillations, and it will end up close to the maximally mixed
state in which both states $|g\rangle$ and $|e\rangle$ have equal
occupation probability. In that case, one could say that
decoherence succeeds in making us lose track of the quantum state
of the qubit but fails to dissipate the energy of the qubit, since
more energy will always be available from the driving field.

\subsection{Combined picture}

We now take the three elements presented above and combine them to
obtain a simple intuitive picture of the problem at hand.

Let us for a moment neglect the effects of decoherence and only
consider the case $\omega\approx\omega_0$. The driving field tries
to flip the state of the qubit alone. However, two of the relevant
eigenstates are entangled states, namely $| 2 \rangle$ and $| 3
\rangle$. One can therefore expect that if the width of the Rabi
peak, or in other words the on-resonance Rabi frequency, is much
larger than the energy separation between the states $| 2 \rangle$
and $| 3 \rangle$, the qubit will start oscillating much faster
than the TLS can respond, and the initial dynamics will look
similar to that of the uncoupled system. Only after many
oscillations and a time of the order of $(E_3-E_2)^{-1}$ will one
start to see the effects of the qubit-TLS interaction. If, on the
other hand, the Rabi frequency is much smaller than the energy
separation between the states $| 2 \rangle$ and $| 3 \rangle$, the
driving field can excite at most one of those two states,
depending on the driving frequency. In that case the qubit-TLS
interactions are strong enough that the TLS can follow
adiabatically the time evolution of the qubit. In the intermediate
region, one expects that if the driving frequency is closer to one
of the two transition frequencies $E_2-E_1$ and $E_3-E_1$, beating
behaviour will be seen right from the beginning. If one looks at
the Rabi peak in the frequency domain, e.g. by plotting the
maximum $g\leftrightarrow e$ qubit-state conversion probability as
a function of frequency, the single peak of the weak-coupling
limit separates into two peaks as the qubit-TLS coupling strength
becomes comparable to and exceeds the on-resonance Rabi frequency.

We do not expect weak to moderate levels of decoherence to cause
any qualitative changes in the qubit dynamics other than, for
example, imposing a decaying envelope on the qubit excitation
probability. As mentioned above, features that are narrower (in
frequency) than the decoherence rates will be suppressed the most.
Note that if the TLS decoherence rates are large enough
\cite{Ashhab}, the TLS can be neglected and one recovers the
single Rabi peak with a height determined by the qubit decoherence
rates alone.

\section{Numerical results}

In the absence of decoherence, we find it easiest to treat the
problem at hand using the dressed-state picture
\cite{Cohen_Tannoudji}. In that picture one thinks of the driving
field mode as being quantized, and processes are described as
involving the absorption and emission of quantized photons by the
qubit+TLS combined system. As a representative case, which also
happens to be the case of most interest to us, we take the
frequency of the driving field to be close to the qubit and TLS
energy splittings. For simplicity, we take those to be equal. We
shall come back to the general case later in this section. Without
going over the rather simple details of the derivation, we show
the four relevant energy levels and the possible transitions in
Fig. 1. The effective Hamiltonian describing the dynamics within
those four levels is given by:

\begin{widetext}
\begin{equation}
\hat{H}_{\rm eff} = \left(
\begin{array}{cccc}
0 & \Omega_0' & \Omega_0' & 0 \\
\Omega_0' & - \delta \omega + \lambda_{\rm cc} -
\lambda_{\rm ss}/2 & 0 & \Omega_0' \\
\Omega_0' & 0 & - \delta \omega + \lambda_{\rm cc} + \lambda_{\rm ss}/2
& - \Omega_0' \\
0 & \Omega_0' & - \Omega_0' & - 2 \delta \omega
\end{array}
\right)
\end{equation}
\end{widetext}

\begin{figure}[ht]
\includegraphics[width=8.5cm]{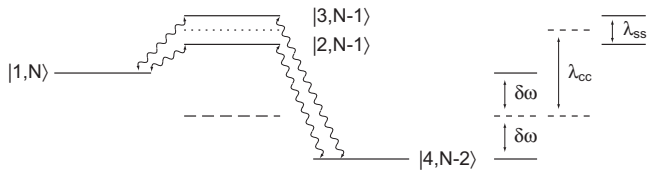}
\caption{\label{fig1} Left: Energy levels and direct transitions
between them in the dressed-state picture. Right: Energy
splittings separated according to their physical origin.}
\end{figure}

\noindent where $\Omega_0'=\Omega_0/2^{3/2}$, $\Omega_0$ is the
on-resonance Rabi frequency in the absence of qubit-TLS coupling,
$\delta \omega=\omega-\omega_0$, and the Hamiltonian is expressed
in the basis of states $(|1,N\rangle, |2,N-1\rangle,
|3,N-1\rangle, |4,N-2\rangle)$, where $N$ is the number of photons
in the driving field. We take the low temperature limit, which
means that we can take the initial state to be $|1,N\rangle$
without the need for any extra initialization. We can now evolve
the system numerically and analyze the dynamics. After we find the
density matrix of the combined qubit+TLS system as a function of
time, we can look at the dynamics of the combined system or that
of the two subsystems separately, depending on which one provides
more insightful information.

We start by demonstrating the separation of the Rabi peak into two
peaks as the qubit-TLS coupling strength is increased. As a
quantifier of the amplitude of Rabi oscillations, we use the
maximum probability for the qubit to be found in the excited state
between times $t=0$ and $t=20\pi/\Omega_0$, and we refer to that
quantity as $P_{\uparrow,{\rm max}}^{\rm (q)}$. In Fig. 2 we plot
$P_{\uparrow,{\rm max}}^{\rm (q)}$ as a function of renormalized
detuning $\delta\omega/\Omega_0$. As was explained in Sec. III,
the peak separates into two when the qubit-TLS coupling strength
exceeds the on-resonance Rabi frequency, up to simple factors of
order one. The system also behaves according to the explanation
given in Sec. III in the weak- and strong-coupling limits. When
$\lambda$ is substantially smaller than $\Omega_0$, oscillations
in the qubit state occur on a time scale $\Omega^{-1}$, where
$\Omega$ is the Rabi frequency defined in Sec. III-B, whereas the
beating behaviour occurs on a time scale $(E_3-E_2)^{-1}$. When
$\lambda$ is more than an order of magnitude smaller than
$\Omega_0$, the effects of the TLS are hardly visible in the qubit
dynamics within the time given above. On the other hand, when
$\lambda$ is large enough such that the energy difference
$E_3-E_2$ is several times larger than $\Omega_0$, the dynamics
corresponds to exciting at most one of the two eigenstates
$|2\rangle$ and $|3\rangle$. We generally see that beating
behaviour becomes less pronounced when the driving frequency is
equal to the qubit energy splitting including the TLS-mean field
shift, i.e. when $\omega=\sqrt{\Delta_{\rm q}^2+(\epsilon_{\rm
q}+\lambda_{\rm cc})^2}$, which corresponds to the top of the
unsplit single peak or the midpoint between the two separated
peaks.

\begin{figure}[ht]
\begin{minipage}[b]{4.2cm}
\includegraphics[width=4.1cm]{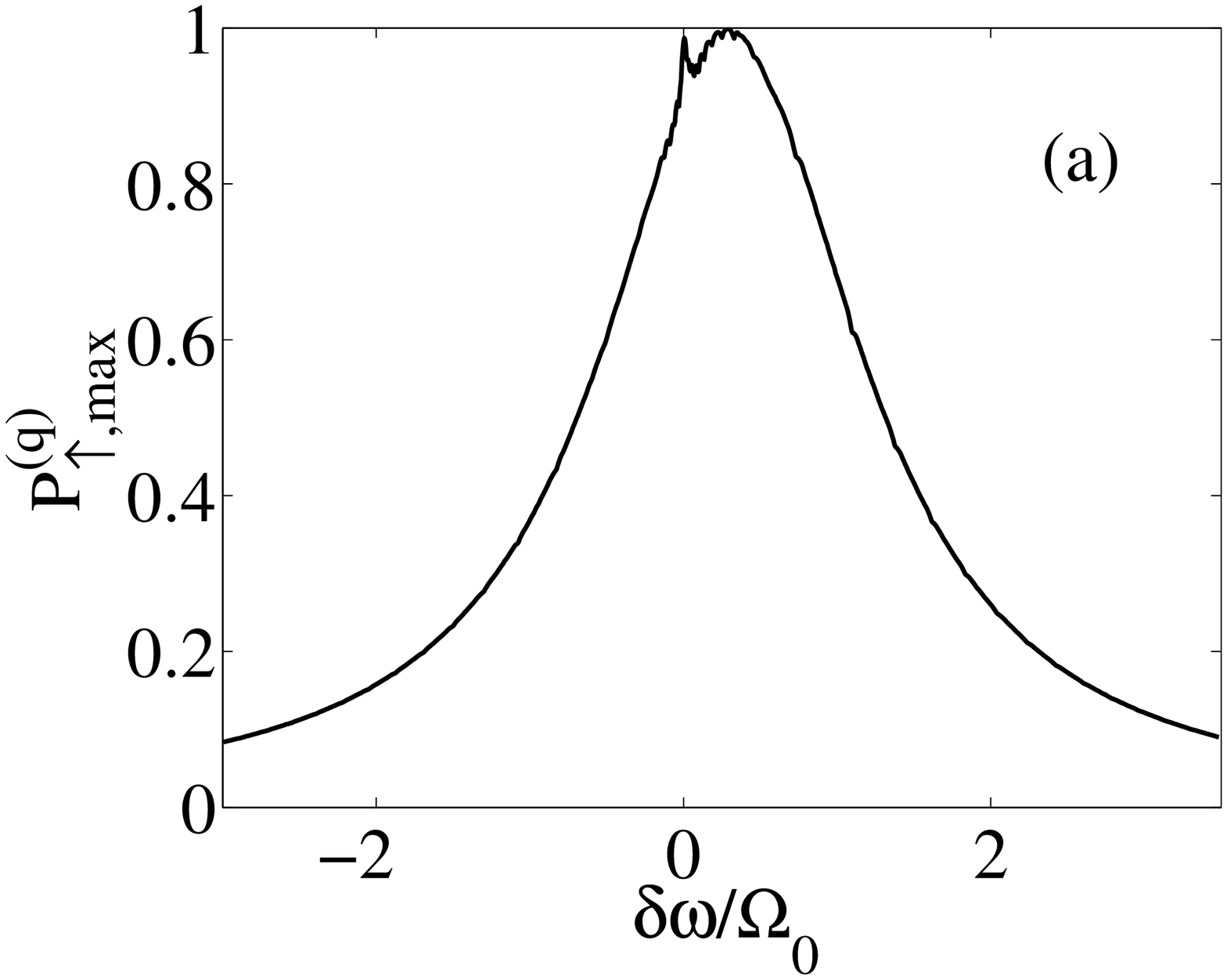}
\end{minipage}
\begin{minipage}[b]{4.2cm}
\includegraphics[width=4.1cm]{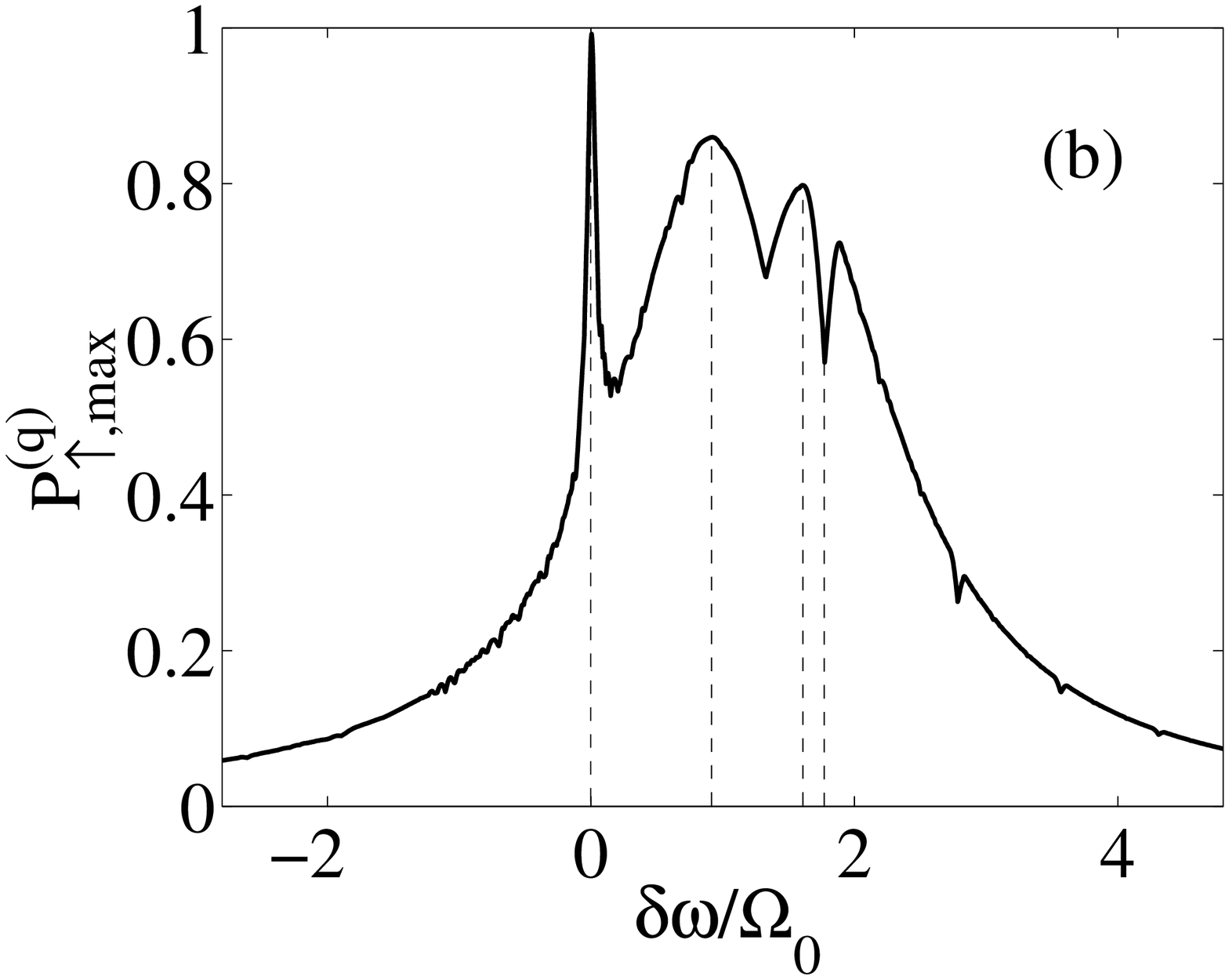}
\end{minipage}
\begin{minipage}[b]{4.2cm}
\includegraphics[width=4.15cm]{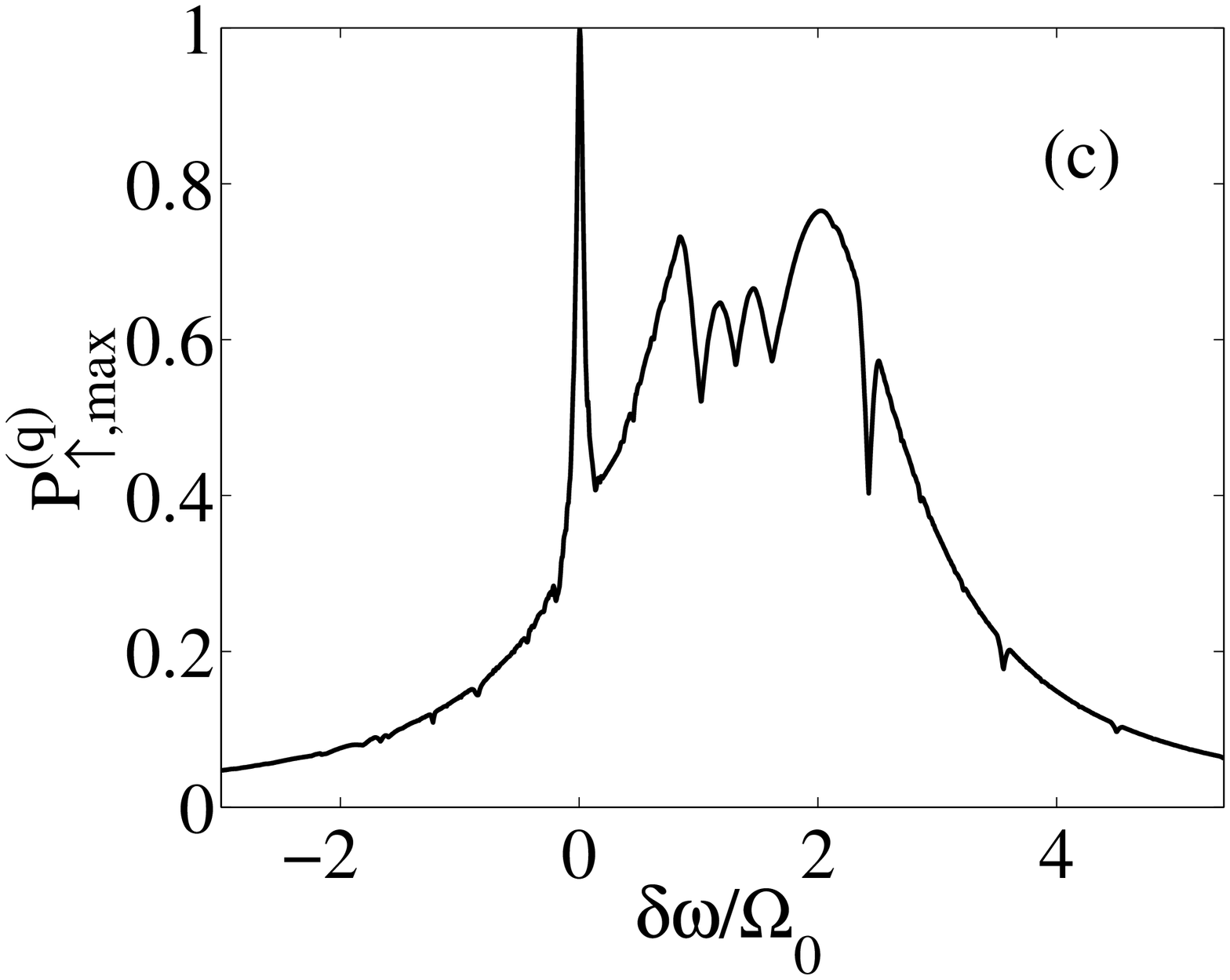}
\end{minipage}
\begin{minipage}[b]{4.2cm}
\includegraphics[width=4.15cm]{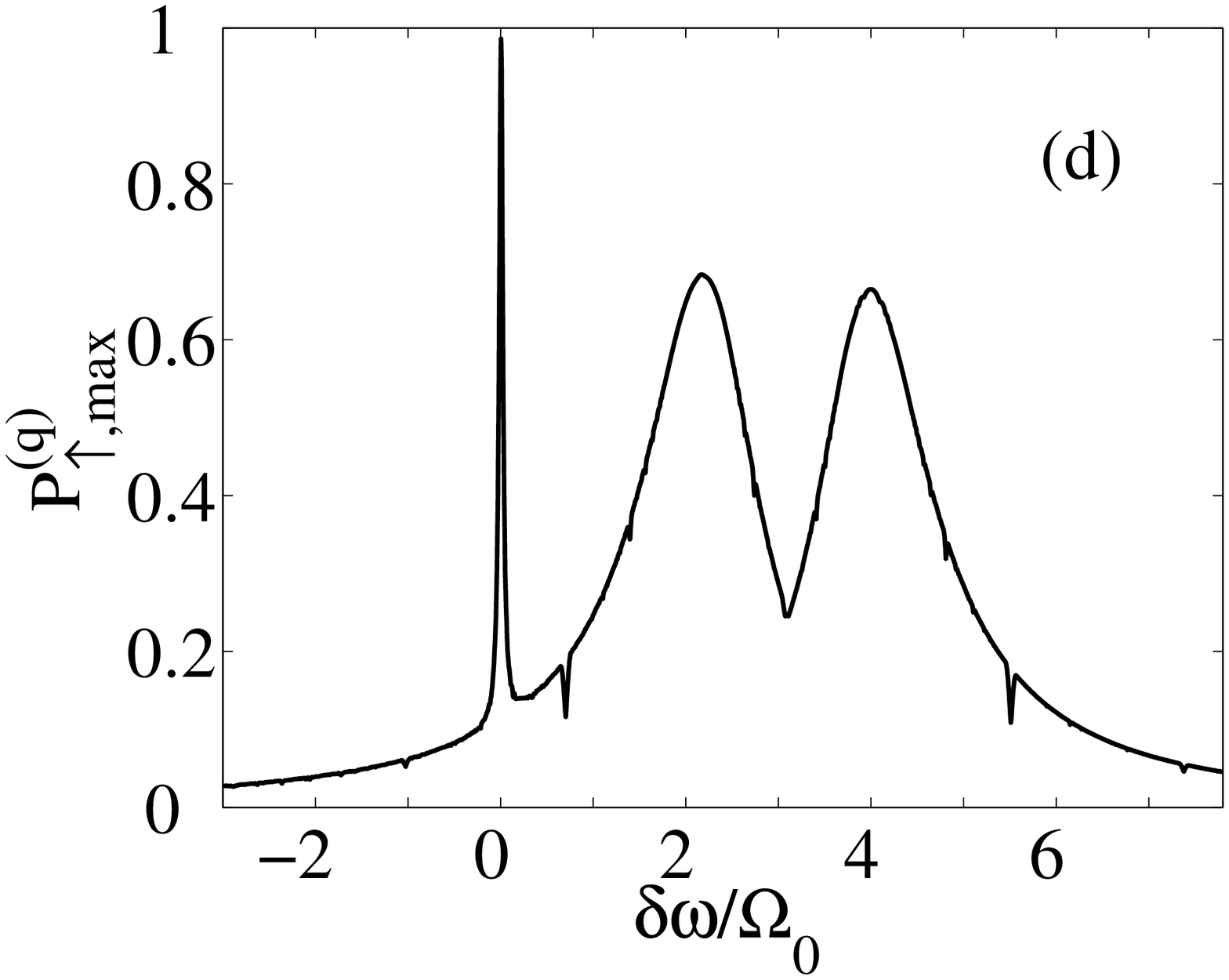}
\end{minipage}
\caption{Maximum qubit excitation probability P$_{\uparrow,{\rm
max}}^{(\rm q)}$ between $t=0$ and $t = 20\pi/\Omega_0$ for
$\lambda/\Omega_0 = $ 0.5 (a), 2 (b), 2.5 (c) and 5 (d).
$\theta_{\rm q}=\pi/4$, and $\theta_{\rm TLS}=\pi/6$.}
\end{figure}

We also see some interesting features in the peak structure of
Fig. 2 that were not discussed in Sec. III. In the
intermediate-coupling regime (Figs. 2b,c), we see a peak that
reaches unit height, i.e. a peak that corresponds to full
$g\leftrightarrow e$ conversion in the qubit dynamics at $\delta
\omega=0$. The asymmetry between the two main peaks in Fig. 2, as
well as the additional dips in the double-peak structure, were
also not immediately obvious from the simple arguments of Sec.
III. In order to give a first explanation of the above features,
we plot in Fig. 3 a curve similar to that in Fig. 2(b) (with
different $\theta_{\rm TLS}$), along with the same quantity
plotted when the eigenstate $|4,N-2\rangle$ is neglected, i.e. by
using a reduced $3\times 3$ Hamiltonian where the fourth row and
column are removed from $\hat{H}_{\rm eff}$. In the three-state
calculation, there is no $\delta\omega=0$ peak, the two main peaks
are symmetric, but we still see some dips. We also plot in Fig. 4
the qubit excitation probability as a function of time for the
four frequencies marked by vertical dashed lines in Fig. 2(b).

\begin{figure}[ht]
\includegraphics[width=8.6cm]{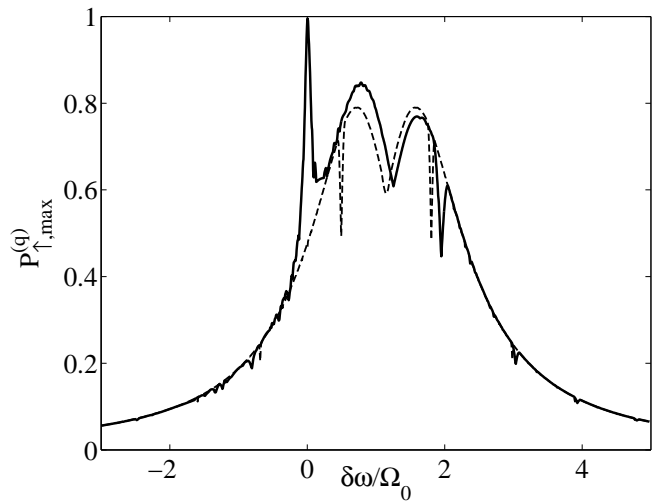}
\caption{Maximum qubit excitation probability P$_{\uparrow,{\rm
max}}^{(\rm q)}$ between $t=0$ and $t = 20\pi/\Omega_0$ for the
four-level system (solid line) and the reduced three-level system
(dashed line). $\lambda/\Omega_0 = 2$, $\theta_{\rm q}=\pi/4$, and
$\theta_{\rm TLS}=\pi/5$.}
\end{figure}

By looking at Fig. 1, one might say that the $\delta\omega=0$ peak
clearly corresponds to a two-photon process coupling states
$|1\rangle$ and $|4\rangle$. In fact, for further demonstration
that this is the case, we have included in Fig. 4 the probability
of the combined qubit+TLS system to be in state $|4\rangle$. This
peak is easiest to observe in the intermediate coupling regime. In
the weak-coupling limit, the qubit and TLS are essentially
decoupled, especially on the time scale of qubit dynamics. In the
strong coupling limit, one can argue that a Raman transition will
give rise to that peak. However, noting that the width of that
peak is of the order of the smaller of the values
$\Omega_0^2/\lambda_{\rm ss}$ and $\Omega_0^2/\lambda_{\rm cc}$,
we can see that it becomes increasingly narrow in that limit. In
other words, the virtual intermediate state after the absorption
of one photon is far enough in energy from the states $|2\rangle$
and $|3\rangle$ to make the peak invisibly narrow. It is rather
surprising, however, that in the intermediate-coupling regime the
peak reaches unit $g\leftrightarrow e$ conversion probability,
even though the transitions to states $|2\rangle$ and $|3\rangle$
are real, rather than being virtual transitions whose role is
merely to mediate the coupling between states $|1\rangle$ and
$|4\rangle$. We have verified that the (almost) unit height of the
peak is quite robust against changes in the angles $\theta_{\rm
q}$ and $\theta_{\rm TLS}$ for a wide range in $\lambda$, even
when that peak coincides with the top of one of the two main
peaks. In fact, the Hamiltonian $\hat{H}_{\rm eff}$ can be
diagonalized rather straightforwardly in the case
$\delta\omega=0$, and one can see that there is no symmetry that
requires full conversion between the states $|1,N\rangle$ and
$|4,N-2\rangle$. The lack of any special relations between the
energy differences in the eigenvalues of $\hat{H}_{\rm eff}$,
however, suggests that almost full conversion should be achieved
in a reasonable amount of time.

\begin{figure}[ht]
\begin{minipage}[b]{4.2cm}
\includegraphics[width=4.2cm]{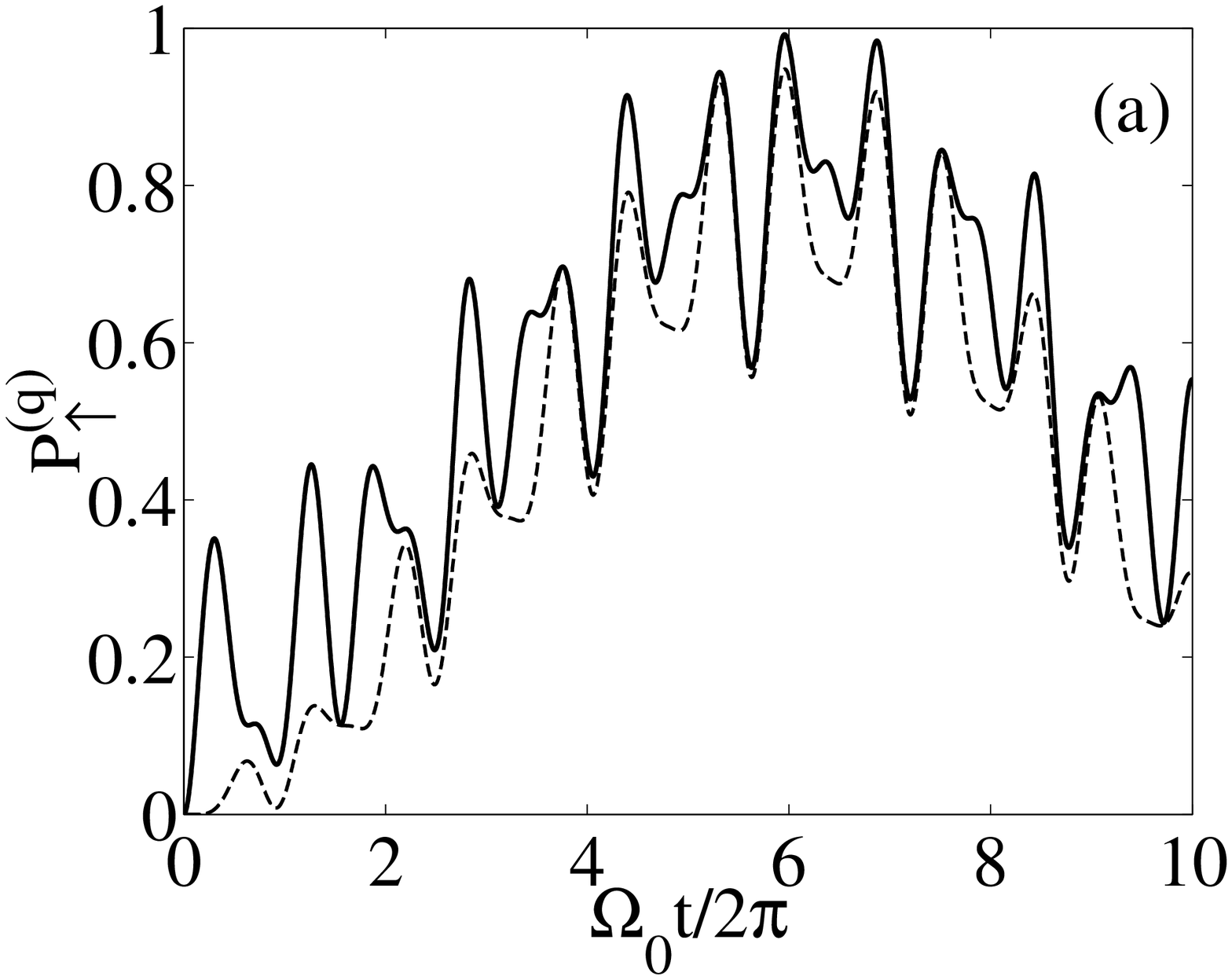}
\end{minipage}
\begin{minipage}[b]{4.2cm}
\includegraphics[width=4.2cm]{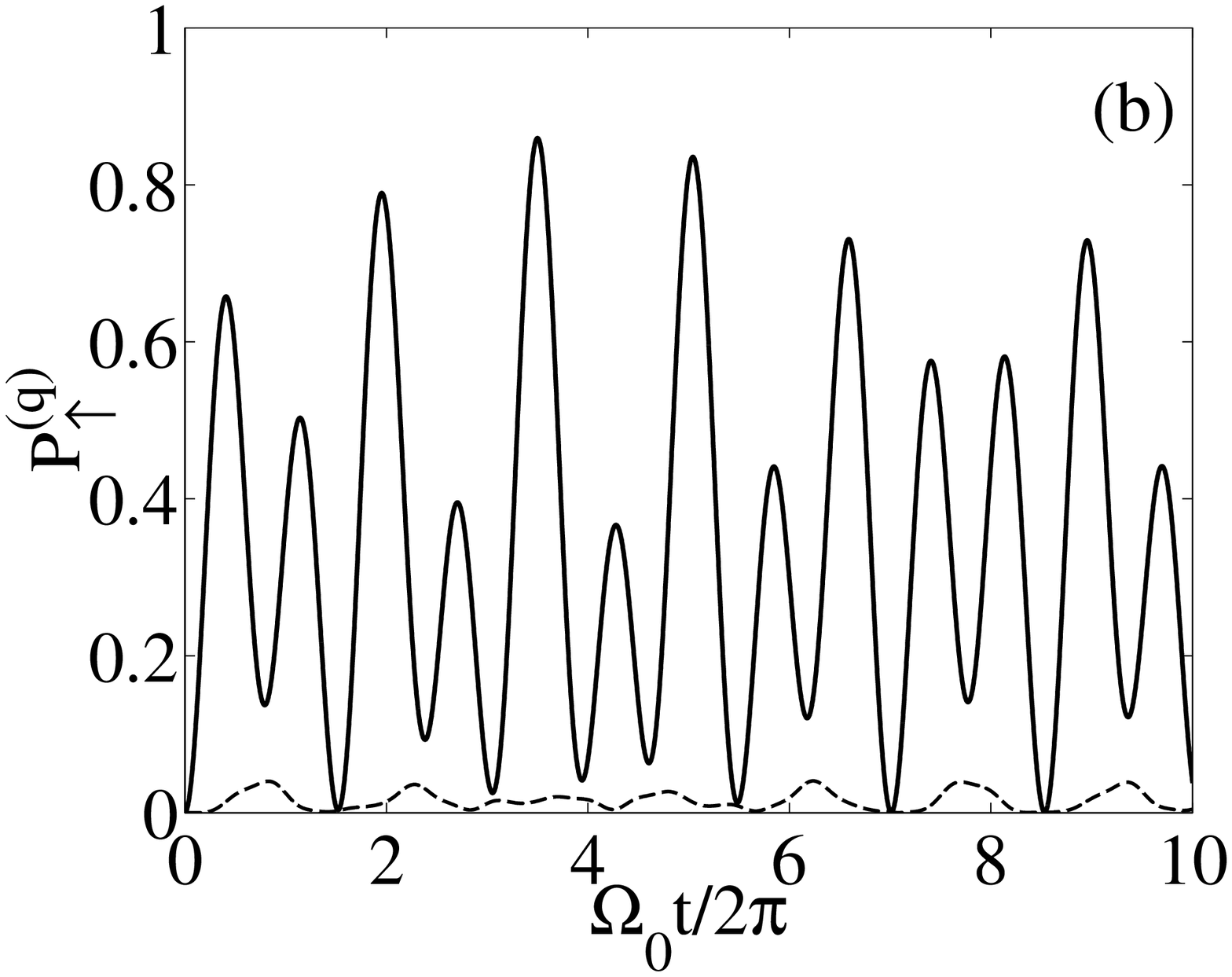}
\end{minipage}
\vspace{0.3cm}
\begin{minipage}[b]{4.2cm}
\includegraphics[width=4.2cm]{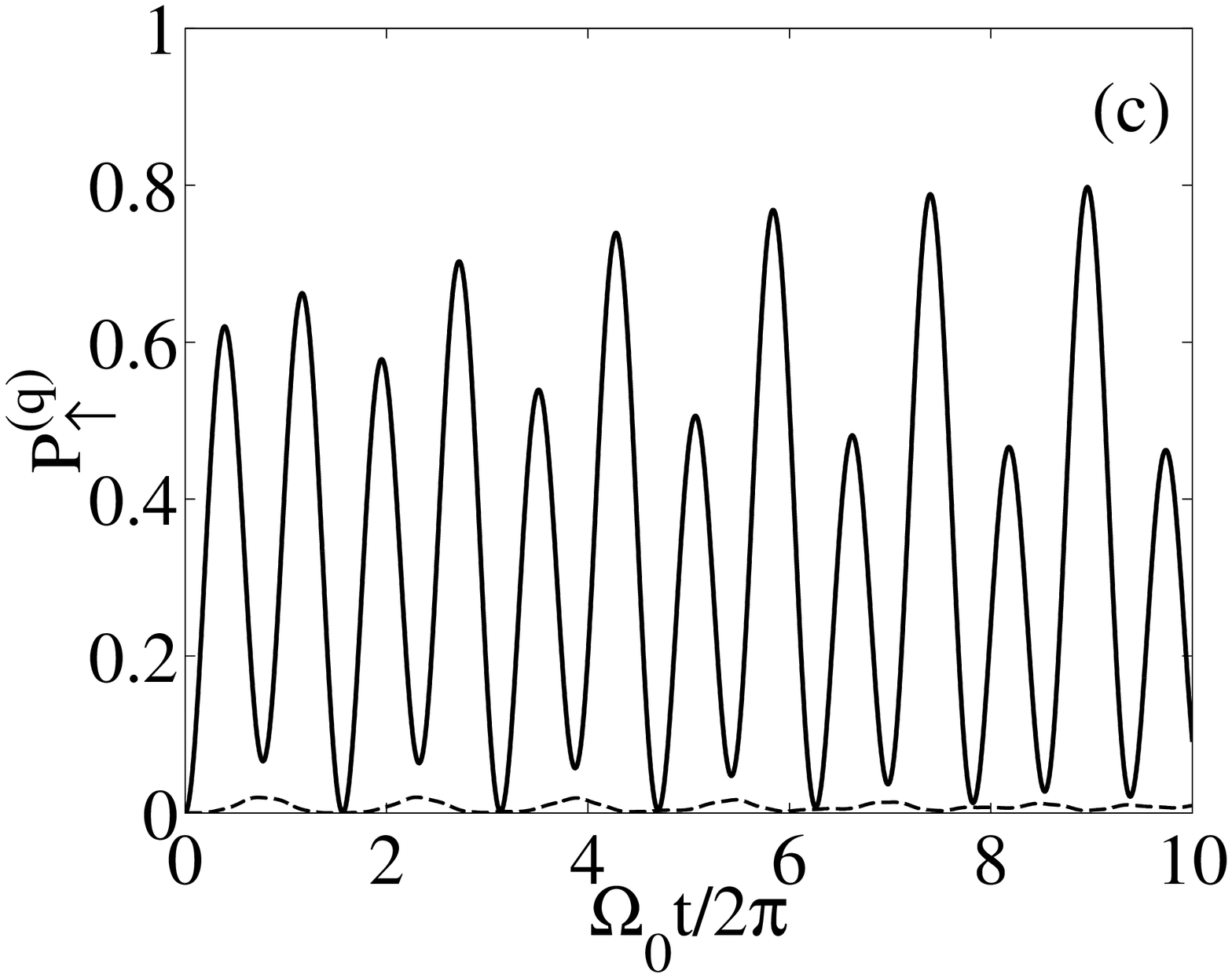}
\end{minipage}
\begin{minipage}[b]{4.2cm}
\includegraphics[width=4.2cm]{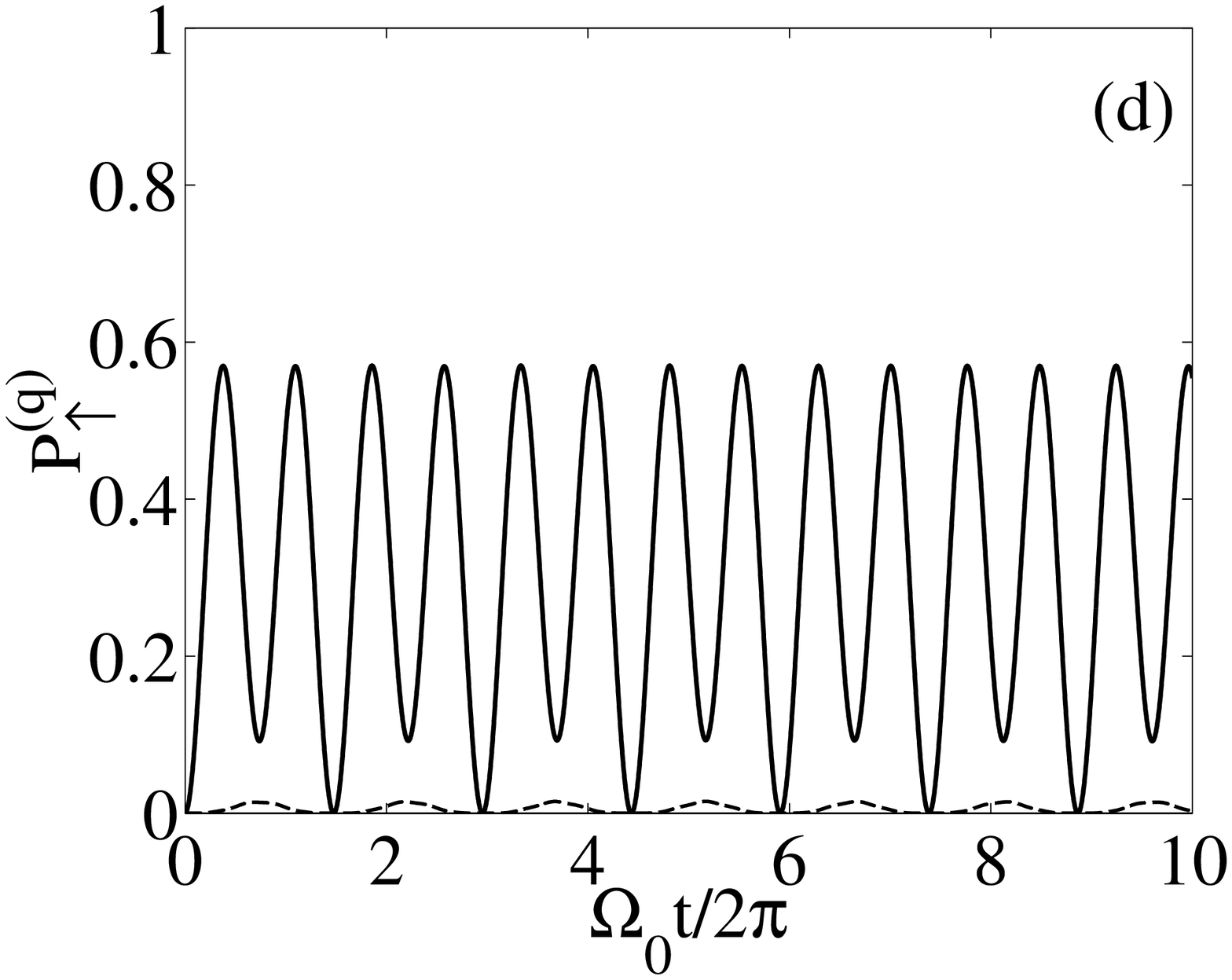}
\end{minipage}
\caption{Qubit excitation probability P$_{\uparrow}^{(\rm q)}$ as
a function of time (solid line) for $\delta\omega/\Omega_0 = $ 0
(a), 0.92 (b), 1.61 (c) and 1.77 (d). The dashed line is the
occupation probability of state $\left|4,N-2\right>$.
$\lambda/\Omega_0 = 2$, $\theta_{\rm q}=\pi/4$, and $\theta_{\rm
TLS}=\pi/5$.}
\end{figure}

The asymmetry between the two main peaks in Figs. 2 and 3 can also
be explained by the fact that in one of those peaks state
$|4\rangle$ is also involved in the dynamics and it increases the
quantity $P_{\uparrow,{\rm max}}^{\rm (q)}$. As above, we have
included in Figs. 4(b) and 4(c) the probability of the combined
qubit+TLS system to be in state $|4\rangle$.

In order to explain the dips in Figs. 2 and 3, we note that the
plotted quantity, $P_{\uparrow,{\rm max}}^{\rm (q)}$, is the sum
of four terms (in the reduced three-level system): a constant and
three oscillating terms. The frequencies of those terms correspond
to the energy differences in the diagonalized $3\times 3$
Hamiltonian. The dips occur at frequencies where the two largest
frequencies are integer multiples of the smallest one. Away from
any such point, $P_{\uparrow,{\rm max}}^{\rm (q)}$ will reach a
value equal to the sum of the amplitudes of the four terms.
Exactly at those points, however, such a constructive buildup of
amplitudes is not always possible, and a dip is generally
obtained. The width of that dip decreases and vanishes
asymptotically as we increase the simulation time, although the
depth remains unaffected.

We also studied the case where the qubit and TLS energy splittings
were different. As can be expected, the effects of the TLS
decrease as it moves away from resonance with the qubit. That is
most clearly reflected in the two-peak structure, where one of the
two main peaks becomes substantially smaller than the other. The
two-photon peak was still clearly observable in plots
corresponding to the same quantity plotted in Fig. 2, i.e. plots
of $P_{\uparrow,{\rm max}}^{\rm (q)}$ vs. $\delta\omega/\Omega_0$,
even when the detuning between the qubit and the TLS was a few
times larger than the coupling strength and the on-resonance Rabi
frequency.

\begin{figure}[ht]
\includegraphics[width=8.2cm]{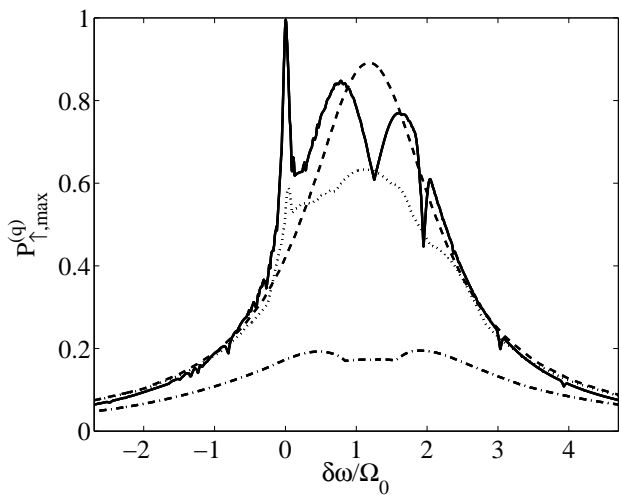}
\caption{Maximum qubit excitation probability P$_{\uparrow,{\rm
max}}^{(\rm q)}$ between $t=0$ and $t=20\pi/\Omega_0$. The solid
line corresponds to the case of no decoherence. The dotted
($\Gamma_{1,2}^{\rm (q)}=0.1\Omega_0/2\pi$ and $\Gamma_{1,2}^{\rm
(TLS)}=0.2\Omega_0/2\pi$), dashed ($\Gamma_{1,2}^{\rm (q)}=0$ and
$\Gamma_{1,2}^{\rm (TLS)}=2\Omega_0$) and dash-dotted
($\Gamma_{1,2}^{\rm (q)}=\Omega_0$ and $\Gamma_{1,2}^{\rm
(TLS)}=0$) lines correspond to different decoherence regimes.
$\lambda/\Omega_0 = 2$, $\theta_{\rm q}=\pi/4$, and $\theta_{\rm
TLS}=\pi/5$. $\Gamma_1^{(\alpha)}$ and $\Gamma_2^{(\alpha)}$ are
the relaxation and dephasing rates of the subsystem $\alpha$,
respectively.}
\end{figure}

The truncated dressed-state picture with four energy levels is
insufficient to study the effects of decoherence. For example,
relaxation from state $|2\rangle$ to $|1\rangle$ does not
necessarily have to involve emission of a photon into the
driving-field mode. We therefore study the effects of decoherence
by treating the driving field classically. We then solve a
Bloch-Redfield master equation with a time-dependent Hamiltonian
and externally-imposed dephasing and relaxation times, as was done
in Ref. \cite{Ashhab}. In Fig. 5 we reproduce the four-level
results of Fig. 3, i.e. $P_{\uparrow,{\rm max}}^{\rm (q)}$ vs.
$\delta\omega/\Omega_0$ with no decoherence, along with the same
quantity obtained when we take into account the effects of
decoherence. For a moderate level of decoherence, we see that the
qubit excitation probability is somewhat reduced and all the
features that are narrower than the decoherence rates are
suppressed partially or completely by the effects of decoherence.
For large qubit decoherence rates, the qubit excitation
probability is greatly reduced close to resonance, where the Rabi
frequency $\Omega$ takes its lowest values. The shallow dip in the
dash-dotted line in Fig. (5) occurs because for those frequencies
and in the absence of decoherence the maximum amplitude is only
reached after several oscillations, whereas it is reached during
the first few oscillations outside that region. For large TLS
decoherence rates, the TLS becomes weakly coupled to the qubit,
and a single peak is recovered in the qubit dynamics (with a
height larger than either the two split peaks). All of these
effects are in agreement with the simple picture presented in Sec.
III.

\section{Experimental considerations}

In the early experiments on phase qubits coupled to TLSs
\cite{Simmonds,Cooper}, the qubit relaxation rate $\Gamma_1^{\rm
(q)}$ ($\sim$40MHz) was comparable to the splitting between the
two Rabi peaks $\lambda_{\rm ss}$ ($\sim$20-70MHz), whereas the
on-resonance Rabi frequency $\Omega_0$ was tunable from 30MHz to
400MHz (note that, as discussed in Sec. III, the Rabi frequency
cannot be reduced to values much lower than the decoherence rates,
or Rabi oscillations would disappear altogether). The large
relaxation rates in those experiments would make several effects
discussed in this paper unobservable. The constraint that
$\Omega_0$ could not be reduced below 30MHz made the
strong-coupling regime, where $\Omega_0\ll\lambda_{\rm ss}$,
inaccessible. The weak-coupling regime, where
$\Omega_0\gg\lambda_{\rm ss}$, was easily accessible in those
experiments. However, as can be seen from Fig. 2, it shows only a
minor signature of the TLS. Although the intermediate-coupling
regime was also accessible, as evidenced by the observation of the
splitting of the Rabi peak into two peaks, observation of the
two-photon process and the additional dips of Fig. 2 discussed
above would have required a time at least comparable to the qubit
relaxation time. That would have made them difficult to
distinguish from experimental fluctuations.

With the new qubit design of Ref. \cite{Martinis2}, the qubit
relaxation time has been increased by a factor of 20. The
constraint that $\Omega_0$ must be at least comparable to
$\Gamma_{1,2}^{\rm (q)}$ no longer prevents accessibility of the
strong-coupling regime. Furthermore, since our simulations were
run for a period of time corresponding to approximately ten Rabi
oscillation cycles, i.e. shorter than the relaxation time observed
in that experiment, all the effects that were discussed above
should be observable, including the observation of the two-photon
peak and the transition from the weak- to the strong-coupling
regimes by varying the driving amplitude.

We finally consider one possible application of our results to
experiments on phase qubits, namely the problem of characterizing
the environment composed of TLSs. As we shall show shortly,
characterizing the TLS parameters and the nature of the qubit-TLS
coupling are not independent questions. The energy splitting of a
given TLS, $E_{\rm TLS}$, can be obtained easily from the location
of the qubit-TLS resonance as the qubit energy splitting is
varied. One can then obtain the distribution of values of $E_{\rm
TLS}$ for a large number of TLSs, as was in fact done in Ref.
\cite{Martinis2}. The splitting of the Rabi resonance peak into
two peaks by itself, however, is insufficient to determine the
values of $\Delta_{\rm TLS}$ and $\epsilon_{\rm TLS}$ separately.
By observing the location of the two-photon peak, in addition to
the locations of the two main peaks, one would be able to
determine both $\lambda_{\rm cc}$ and $\lambda_{\rm ss}$ for a
given TLS, as can be seen from Fig. 1. Those values can then be
used to calculate both $E_{\rm TLS}$ and $\theta_{\rm TLS}\equiv
\arctan (\Delta_{\rm TLS}/\epsilon_{\rm TLS})$ of that TLS. The
distribution of values of $\theta_{\rm TLS}$ can then be used to
test models of the environment, such as the one given in Ref.
\cite{Shnirman} to describe the results of Ref. \cite{Astafiev}.

In order to reach the above conclusion, we have made the
assumption that the distribution of values of $\theta_{\rm TLS}$
for those TLSs with sufficiently strong coupling to the qubit is
representative of all TLSs. Since it is generally believed that
strong coupling is a result of proximity to the junction, the
above assumption is quite plausible, as long as the other TLSs
share the same nature. Although it is possible that there might be
two different types of TLSs of different nature in a qubit's
environment, identifying that possibility would also be helpful in
understanding the nature of the environment. We have also assumed
that $\theta_{\rm q}$ does not take the special value $\pi/2$
(note that, based on the arguments of Refs.
\cite{Shnirman,Astafiev}, we are also assuming that generally
$\theta_{\rm TLS}\neq\pi/2$). That assumption would not raise any
concern when dealing with charge or flux qubits, where both
$\Delta_{\rm q}$ and $\epsilon_{\rm q}$ can be adjusted in a
single experiment, provided an appropriate design is used.
However, the situation is trickier with phase qubits. The results
in that case depend on the nature of the qubit-TLS coupling, which
we discuss next.

The two mechanisms that are currently considered the most likely
candidates to describe the qubit-TLS coupling are through either
(1) a dependence of the Josephson junction's critical current on
the TLS state or (2) Coulomb interactions between a charged TLS
and the charge across the junction. In the former case, one has an
effective value of $\theta_{\rm q}$ that is different from $\pi/2$
\cite{Zagoskin}, and the assumption of an intermediate value of
$\theta_{\rm q}$ is justified. In the case of coupling through
Coulomb interactions, on the other hand, one effectively has
$\theta_{\rm q}=\pi/2$, and therefore $\lambda_{\rm cc}$ vanishes
for all the TLSs. In that case the two-photon peak would always
appear at the midpoint (to a good approximation) between the two
main Rabi peaks. Although that would prevent the determination of
the distribution of values of $\theta_{\rm TLS}$, it would be a
strong indication that Coulomb interactions with the charge across
the junction are responsible for the qubit-TLS coupling rather
than the critical current dependence on the TLS state. Note also
that if it turns out that this is in fact the case, and the
distribution of values of $\theta_{\rm TLS}$ cannot be extracted
from the experimental results, that distribution might be
irrelevant to the question of decoherence in phase qubits.

\section{Conclusion}

We have studied the problem of a harmonically-driven qubit that is
interacting with an uncontrollable two-level system and a
background environment. We have presented a simple picture to
understand the majority of the phenomena that are observed in this
system. That picture is composed of three elements: (1) the
four-level energy spectrum of the qubit+TLS system, (2) the basic
properties of the Rabi-oscillation dynamics and (3) the basic
effects of decoherence. We have confirmed the predictions of that
picture using a systematic numerical analysis where we have varied
a number of relevant parameters. We have also found unexpected
features in the resonance-peak structure. We have analyzed the
behaviour of the system and provided simple explanations in those
cases as well. Our results can be tested with available
experimental systems. Furthermore, they can be used in
experimental attempts to characterize the TLSs surrounding a
qubit, which can then be used as part of possible techniques to
eliminate the TLSs' detrimental effects on the qubit operation.

\begin{acknowledgments}
This work was supported in part by the Army Research Office (ARO),
Laboratory of Physical Sciences (LPS), National Security Agency
(NSA) and Advanced Research and Development Activity (ARDA) under
Air Force Office of Research (AFOSR) contract number
F49620-02-1-0334; and also supported by the National Science
Foundation grant No.~EIA-0130383. One of us (S. A.) was supported
by a fellowship from the Japan Society for the Promotion of
Science (JSPS).
\end{acknowledgments}


\begin{thebibliography}{99}
\bibitem{Assumptions} We shall make a number of assumptions that
might seem too specific at certain points. We have been careful,
however, not to make a choice of parameters that causes any of the
interesting physical phenomena to disappear. For a more detailed
discussion of our assumptions, see \cite{Ashhab}.
\end{thebibliography}
\end{document}